\documentclass[twocolumn,showpacs,amsmath,amssymb,superscriptaddress,
draft,prl]{revtex4}
\usepackage{graphicx}
\usepackage{dcolumn}
\usepackage{bm}

\newcommand{\nc}{\newcommand}
\nc{\rnc}{\renewcommand}
\nc{\nn}{\nonumber}
\nc{\ch}{\cosh}
\nc{\sh}{\sinh}
\rnc{\th}{\tanh}
\nc{\db}{\displaybreak[0]\\}
\nc{\bra}{\langle}
\nc{\ket}{\rangle}
\rnc{\(}{\left(}
\rnc{\)}{\right)}
\nc{\xxx}{$XXX \,$}
\nc{\lam}{\lambda}
\def\grapha{
 \begin{picture}(15,9)(-0.5,6)
 \put(2,2){\circle{1}}
 \put(12,2){\circle{1}}
 \put(2,12){\circle{1}}
 \put(12,12){\circle{1}}

 \put(2.5,2){\vector(1,0){6}}
 \put(8.5,2){\line(1,0){3}}

 \put(11.5,12){\vector(-1,0){6}}
 \put(2.5,12){\line(1,0){3}}

 \put(2,2.5){\line(0,1){9}}
 
 \put(12,2.5){\line(0,1){9}}

 \put(2.353,2.353){\line(1,1){9.3}}
 \put(2.353,11.647){\line(1,-1){9.3}}

 \put(-0.4,0.){{\scriptsize $4$}}
 \put(13,0.){{\scriptsize $3$}}
 \put(-0.4,12.5){{\scriptsize $1$}}
 \put(13,12.5){{\scriptsize $2$}}
 \end{picture}}
\def\graphb{
 \begin{picture}(15,9)(-0.5,6)
 \put(2,2){\circle{1}}
 \put(12,2){\circle{1}}
 \put(2,12){\circle{1}}
 \put(12,12){\circle{1}}

 \put(2,2.5){\line(0,1){9}}
 
 \put(12,2.5){\line(0,1){9}}

 \put(2.353,2.353){\line(1,1){9.3}}
 \put(2.353,11.647){\line(1,-1){9.3}}

 \put(-0.4,0.){{\scriptsize $4$}}
 \put(13,0.){{\scriptsize $3$}}
 \put(-0.4,12.5){{\scriptsize $1$}}
 \put(13,12.5){{\scriptsize $2$}}
 \end{picture}}
\def\graphc{
 \begin{picture}(15,9)(-0.5,6)
 \put(2,2){\circle{1}}
 \put(12,2){\circle{1}}
 \put(2,12){\circle{1}}
 \put(12,12){\circle{1}}

 \put(11.5,12){\vector(-1,0){6}}
 \put(2.5,12){\line(1,0){3}}

 \put(12,2.5){\line(0,1){9}}

 \put(2.353,2.353){\line(1,1){9.3}}

 \put(-0.4,0.){{\scriptsize $4$}}
 \put(13,0.){{\scriptsize $3$}}
 \put(-0.4,12.5){{\scriptsize $1$}}
 \put(13,12.5){{\scriptsize $2$}}
 \end{picture}}
\def\graphd{
 \begin{picture}(15,9)(-0.5,6)
 \put(2,2){\circle{1}}
 \put(12,2){\circle{1}}
 \put(2,12){\circle{1}}
 \put(12,12){\circle{1}}

 \put(11.5,12){\vector(-1,0){6}}
 \put(2.5,12){\line(1,0){3}}

 \put(2,2.5){\line(0,1){9}}
 
 \put(12,2.5){\line(0,1){9}}

 \put(-0.4,0.){{\scriptsize $4$}}
 \put(13,0.){{\scriptsize $3$}}
 \put(-0.4,12.5){{\scriptsize $1$}}
 \put(13,12.5){{\scriptsize $2$}}
 \end{picture}}
\def\graphe{
 \begin{picture}(15,9)(-0.5,6)
 \put(2,2){\circle{1}}
 \put(12,2){\circle{1}}
 \put(2,12){\circle{1}}
 \put(12,12){\circle{1}}

 \put(2,2.5){\line(0,1){9}}
 
 \put(2.353,2.353){\line(1,1){9.3}}

 \put(2.353,11.647){\line(1,-1){9.3}}

 \put(-0.4,0.){{\scriptsize $4$}}
 \put(13,0.){{\scriptsize $3$}}
 \put(-0.4,12.5){{\scriptsize $1$}}
 \put(13,12.5){{\scriptsize $2$}}
 \end{picture}}
\def\graphf{
 \begin{picture}(15,9)(-0.5,6)
 \put(2,2){\circle{1}}
 \put(12,2){\circle{1}}
 \put(2,12){\circle{1}}
 \put(12,12){\circle{1}}

 \put(11.5,12){\vector(-1,0){6}}
 \put(2.5,12){\line(1,0){3}}

 \put(2.353,2.353){\line(1,1){9.3}}

 \put(2.353,11.647){\line(1,-1){9.3}}

 \put(-0.4,0.){{\scriptsize $4$}}
 \put(13,0.){{\scriptsize $3$}}
 \put(-0.4,12.5){{\scriptsize $1$}}
 \put(13,12.5){{\scriptsize $2$}}
 \end{picture}}
\def\graphg{
 \begin{picture}(15,9)(-0.5,6)
 \put(2,2){\circle{1}}
 \put(12,2){\circle{1}}
 \put(2,12){\circle{1}}
 \put(12,12){\circle{1}}

 \put(11.5,12){\vector(-1,0){6}}
 \put(2.5,12){\line(1,0){3}}

 \put(2,2.5){\line(0,1){9}}
 
 \put(2.353,11.647){\line(1,-1){9.3}}

 \put(-0.4,0.){{\scriptsize $4$}}
 \put(13,0.){{\scriptsize $3$}}
 \put(-0.4,12.5){{\scriptsize $1$}}
 \put(13,12.5){{\scriptsize $2$}}
 \end{picture}}
\def\graphh{
 \begin{picture}(15,9)(-0.5,6)
 \put(2,2){\circle{1}}
 \put(12,2){\circle{1}}
 \put(2,12){\circle{1}}
 \put(12,12){\circle{1}}

 \put(2.5,2){\vector(1,0){6}}
 \put(8.5,2){\line(1,0){3}}

 \put(11.5,12){\vector(-1,0){6}}
 \put(2.5,12){\line(1,0){3}}

 \put(12,2.5){\line(0,1){9}}

 \put(-0.4,0.){{\scriptsize $4$}}
 \put(13,0.){{\scriptsize $3$}}
 \put(-0.4,12.5){{\scriptsize $1$}}
 \put(13,12.5){{\scriptsize $2$}}
 \end{picture}}
\def\graphi{
 \begin{picture}(15,9)(-0.5,6)
 \put(2,2){\circle{1}}
 \put(12,2){\circle{1}}
 \put(2,12){\circle{1}}
 \put(12,12){\circle{1}}

 \put(2.5,2){\vector(1,0){6}}
 \put(8.5,2){\line(1,0){3}}

 \put(11.5,12){\vector(-1,0){6}}
 \put(2.5,12){\line(1,0){3}}

 \put(2.353,2.353){\line(1,1){9.3}}

 \put(2.353,11.647){\line(1,-1){9.3}}

 \put(-0.4,0.){{\scriptsize $4$}}
 \put(13,0.){{\scriptsize $3$}}
 \put(-0.4,12.5){{\scriptsize $1$}}
 \put(13,12.5){{\scriptsize $2$}}
 \end{picture}}

\def\graphj{
 \begin{picture}(15,9)(-0.5,6)
 \put(2,2){\circle{1}}
 \put(12,2){\circle{1}}
 \put(2,12){\circle{1}}
 \put(12,12){\circle{1}}

 \put(2,2.5){\line(0,1){9}}
 
 \put(2.353,11.647){\line(1,-1){9.3}}

 \put(-0.4,0.){{\scriptsize $4$}}
 \put(13,0.){{\scriptsize $3$}}
 \put(-0.4,12.5){{\scriptsize $1$}}
 \put(13,12.5){{\scriptsize $2$}}
 \end{picture}}
\def\graphk{
 \begin{picture}(15,9)(-0.5,6)
 \put(2,2){\circle{1}}
 \put(12,2){\circle{1}}
 \put(2,12){\circle{1}}
 \put(12,12){\circle{1}}

 \put(11.5,12){\vector(-1,0){6}}
 \put(2.5,12){\line(1,0){3}}

 \put(2,2.5){\line(0,1){9}}
 
 \put(2.353,11.647){\line(1,-1){9.3}}

 \put(-0.4,0.){{\scriptsize $4$}}
 \put(13,0.){{\scriptsize $3$}}
 \put(-0.4,12.5){{\scriptsize $1$}}
 \put(13,12.5){{\scriptsize $2$}}
 \end{picture}}
\def\graphl{
 \begin{picture}(15,9)(-0.5,6)
 \put(2,2){\circle{1}}
 \put(12,2){\circle{1}}
 \put(2,12){\circle{1}}
 \put(12,12){\circle{1}}

 \put(2.5,12){\line(1,0){9}}

 \put(2,2.5){\line(0,1){9}}
 
 \put(2.353,11.647){\line(1,-1){9.3}}

 \put(-0.4,0.){{\scriptsize $4$}}
 \put(13,0.){{\scriptsize $3$}}
 \put(-0.4,12.5){{\scriptsize $1$}}
 \put(13,12.5){{\scriptsize $2$}}
 \end{picture}}
\def\graphm{
 \begin{picture}(15,9)(-0.5,6)
 \put(2,2){\circle{1}}
 \put(12,2){\circle{1}}
 \put(2,12){\circle{1}}
 \put(12,12){\circle{1}}

 \put(2.5,2){\line(1,0){9}}

 \put(2.5,12){\line(1,0){9}}

 \put(2,2.5){\line(0,1){9}}
 
 \put(12,2.5){\line(0,1){9}}

 \put(-0.4,0.){{\scriptsize $4$}}
 \put(13,0.){{\scriptsize $3$}}
 \put(-0.4,12.5){{\scriptsize $1$}}
 \put(13,12.5){{\scriptsize $2$}}
 \end{picture}}

\def\graphaa{
 \begin{picture}(15,3)(-0.5,1.3)
 \put(2,2){\circle{1}}
 \put(12,2){\circle{1}}
 
 \put(2.5,2){\line(1,0){9}}

 \put(-0.4,0.){{\scriptsize $j$}}
 \put(-0.4,-4){{\scriptsize$k$}}
 \put(13,0.){{\scriptsize $k$}}
 \put(13,-4.){{\scriptsize $j$}}
 \end{picture}}
\def\graphbb{
 \begin{picture}(15,3)(-0.5,1.3)
 \put(2,2){\circle{1}}
 \put(12,2){\circle{1}}
 
 \put(2.5,2){\vector(1,0){6}}
 \put(8.5,2){\line(1,0){3}}

 \put(-0.4,0.){{\scriptsize $j$}}
 \put(13,0.){{\scriptsize $k$}}
 \end{picture}}

\def\graphdd{
 \begin{picture}(15,3)(-0.5,1.3)
 \put(2,2){\circle{1}}
 \put(12,2){\circle{1}}
\put(2.5,2){\line(1,0){9}}
\put(-0.4,0.){{\scriptsize $1$}}
 \put(13,0.){{\scriptsize $2$}}
 \end{picture}}
\def\graphee{
 \begin{picture}(15,3)(-0.5,1.3)
 \put(2,2){\circle{1}}
 \put(12,2){\circle{1}}
\put(2.5,2){\line(1,0){9}}
\put(-0.4,0.){{\scriptsize $4$}}
 \put(13,0.){{\scriptsize $3$}}
 \end{picture}}
\begin{document}
%
\title{Third Neighbor Correlators of Spin-1/2 Heisenberg Antiferromagnet}
%
\author{Kazumitsu Sakai}
\affiliation{Institute for Solid State Physics, University of Tokyo, Kashiwa,
Chiba 277-8581, Japan}
\author{Masahiro Shiroishi}
\affiliation{Institute for Solid State Physics, University of Tokyo, Kashiwa,
Chiba 277-8581, Japan}
\author{Yoshihiro Nishiyama}
\affiliation{Department of Physics, Faculty of Science, Okayama University,
Okayama 700-8530, Japan}
\author{Minoru Takahashi}
\affiliation{Institute for Solid State Physics, University of Tokyo, Kashiwa,
Chiba 277-8581, Japan}
\date{February 27, 2003}
%
%
\begin{abstract}
We exactly evaluate the third neighbor correlator 
$\langle S_j^z S_{j+3}^z \rangle$ and all the
possible non-zero correlators 
$\langle S^{\alpha}_j S^{\beta}_{j+1} S^{\gamma}_{j+2} S^{\delta}_{j+3}
\rangle$ of the spin-1/2 Heisenberg $XXX$ antiferromagnet 
in the ground state without magnetic field.
All the  correlators are expressed in terms of certain 
combinations of logarithm $\ln 2$, the Riemann zeta function 
$\zeta(3)$, $\zeta(5)$ with rational coefficients.
The results accurately coincide  with the numerical
ones obtained by the density-matrix renormalization group
 method and the numerical diagonalization.
\end{abstract}
\maketitle
%
%
Since Bethe's pioneering work 
of the spin-1/2 Heisenberg \xxx 
magnet \cite{Bethe}, 
%
\begin{equation}
H=J\sum_{j=1}^{L}\(S_j^xS_{j+1}^x+S_j^yS_{j+1}^y+S_j^zS_{j+1}^z\),
\label{Hami}
\end{equation} 
%
exact evaluation of the correlation 
functions has been a long standing problem in mathematical 
physics. Especially significant are the spin-spin correlators 
$\bra S_j^z S_{j+k}^z \ket$ (or equivalently 
$\bra S_j^+ S_{j+k}^-\ket/2$; $S_j^{\pm}=S_j^x\pm i S_j^y$), 
for which only the first 
and second neighbor ($k=1,2$) have been calculated so far:
%
\begin{align}
\bra S_j^{z} S_{j+1}^z\ket&=\frac{1}{12}-\frac{1}{3}\ln2
\simeq-0.14771572685, \label{1st} \db
\bra S_j^{z} S_{j+2}^z\ket&=\frac{1}{12}-\frac{4}{3}\ln2+
\frac{3}{4}\zeta(3)\simeq0.06067976996,
\label{2nd}
\end{align}
%
where $\zeta(s)$ is the Riemann zeta function and 
$\bra \cdots \ket$ denotes the ground state expectation value 
of the antiferromagnetic model ($J>0$).
Here we have taken the thermodynamic limit $L\to\infty$. In this 
letter, we would like to report our new results regarding the 
third neighbor correlators. Our main result is
%
\begin{align}
\bra S^z_j & S^z_{j+3}\ket=\frac{1}{2}\bra S_j^+S_{j+3}^-\ket \nn \db
&=\frac{1}{12}-
3\ln 2+\frac{37}{6}\zeta(3)
-\frac{14}{3}\zeta(3)\ln 2 
-\frac{3}{2}\zeta(3)^2 \nn \db
&\quad-\frac{125}{24}\zeta(5)+\frac{25}{3}\zeta(5)\ln 2
\simeq -0.05024862726.
\label{3rd}
\end{align}
%
In addition, we obtain the 
third neighbor one-particle Green function
$\bra c^{\dagger}_j c_{j+3} \ket_{\rm f}$
%
\begin{align}
&\bra c^{\dagger}_j c_{j+3} \ket_{\rm f}=
\frac{1}{30}-2\ln 2+\frac{169}{30}\zeta(3)-\frac{10}{3}\zeta(3)\ln 2 
\nn \db
&-\frac{6}{5}\zeta(3)^2-\frac{65}{12}\zeta(5)+
\frac{20}{3}\zeta(5)\ln 2
\simeq 0.08228771669,
\label{green}
\end{align}
for the isotropic spinless fermion model 
corresponding to \eqref{Hami} by the Jordan-Wigner 
transformation:
\begin{equation}
S_k^{-}=\prod_{j=1}^{k-1}(1-2c_j^{\dagger}c_j)
c_k^{\dagger},\quad
S_k^+=\prod_{j=1}^{k-1}(1-2c_j^{\dagger}c_j)c_k.
\label{JW}
\end{equation}
Here $\bra\dots\ket_{\rm f}$ denotes
the expectation value in the half-filled state
of the spinless fermion model.
Moreover we exactly calculate  all
the possible non-zero correlators $\langle S^{\alpha}_j 
S^{\beta}_{j+1} S^{\gamma}_{j+2}S^{\delta}_{j+3} \rangle$.

The result \eqref{1st} comes from the ground
state energy of \eqref{Hami} derived by Hulth\'{e}n in 
1938 \cite{hult}. The result \eqref{2nd} was obtained 
by one of the authors in 1977 \cite{Taka,Takabook} via 
the strong coupling expansion for 
the ground state energy of the half-filled Hubbard model.
This result is also reproduced in the framework of the 
asymptotic Bethe ansatz for an integrable spin chain with 
variable range exchange \cite{DitIno}. However, probably due 
to the complexity of the wave function for these models,
no one has succeeded in generalizing the method to
obtain the higher neighbor correlators.

On the other hand, utilizing the representation theory of 
the quantum affine algebra $U_q(\widehat{sl_2})$ and the 
associated vertex operators, in 1992, Jimbo \textit{et al.} 
derived a universal 
multiple integral representation  of arbitrary correlators 
for the massive $XXZ$ antiferromagnet \cite{JMMN,JMbook}. 
Their result has been extended to the \xxx 
\cite{Naka,KIEU}, the massless $XXZ$ \cite{JM,KMT} and 
the $XYZ$ \cite{Quano} antiferromagnets. 
However the explicit evaluation, even for the second neighbor 
correlator \eqref{2nd}, was not achieved for a long time.

In this respect, it is quite remarkable that Boos and Korepin 
recently devised a general method to evaluate the
multiple integral representation especially
in the study of the Emptiness Formation Probability (EFP) 
for the \xxx antiferromagnet \cite{BK1,BK2}. 
The EFP, $P(n)$ describes the probability of finding a 
ferromagnetic string of length $n$ in the antiferromagnetic 
ground state \cite{KIEU}. Explicitly it reads
\begin{equation}
P(n)=\biggl\bra \prod_{j=1}^n \(S_j^z+\frac{1}{2}\)\biggr\ket.
\label{efp}
\end{equation}
By reducing the integrand of the multiple integral representation 
to certain \textit{canonical form}, the EFP 
for $n=3,4$ \cite{BK1,BK2} and $n=5$ \cite{BKNS} were
evaluated by Boos \textit{et al.}
(see also recent progress for $n=6$ \cite{BKS}). 
Note that  $P(2)$ and $P(3)$ in \eqref{efp} are related 
to the first and second neighbor correlators as
$P(2)=1/4+\bra S_j^zS_{j+1}^z\ket$ and
$P(3)=1/8+\bra S_j^zS_{j+1}^z\ket+\bra S_j^zS_{j+2}^z\ket/2$.

Here we quote the explicit form of $P(4)$ obtained in
\cite{BK1,BK2}, which is closely related to the
third neighbor correlator $\bra S_j^z S_{j+3}^z\ket$.
%
\begin{align}
P(4)=&\frac{1}{16}+\frac{3}{4}\bra S_j^z S_{j+1}^z\ket+
\frac{1}{2}\bra S_j^z S_{j+2}^z\ket
+\frac{1}{4}\bra S_{j}^zS_{j+3}^z\ket \nn \db
&+\bra S_j^z S_{j+1}^zS_{j+2}^zS_{j+3}^z \ket \nn \db
=&\frac{1}{5}-2\ln2+\frac{173}{60}\zeta(3)-
\frac{11}{6}\zeta(3)\ln2-\frac{51}{80}\zeta(3)^2 \nn \db
&-\frac{55}{24}\zeta(5)+\frac{85}{24}\zeta(5)\ln 2.
\label{p4}
\end{align}
%
Note that on the antiferromagnetic ground state without magnetic
field, all the correlators 
with an odd number of $S^z$ vanishes. Substituting \eqref{1st} and
\eqref{2nd} into \eqref{p4}, one finds the relation between the 
third neighbor correlator $\bra S_{j}^zS_{j+3}^z\ket$ and the 
four point correlator 
$\bra S_j^z S_{j+1}^zS_{j+2}^zS_{j+3}^z \ket$. However the
exact value of 
$\bra S_j^{z}S_{j+3}^z\ket$ itself
can not be determined solely from $P(4)$.

To determine $\bra S_j^z S_{j+3}^z\ket$, 
we consider the following auxiliary correlator:
%
\begin{align}
P_{+-+-}^{+-+-}
%
=&\frac{1}{16}-\frac{3}{4}\bra S_j^z S_{j+1}^z\ket+
\frac{1}{2}\bra S_j^z S_{j+2}^z\ket
-\frac{1}{4}\bra S_{j}^zS_{j+3}^z\ket \nn \db
&+\bra S_j^z S_{j+1}^zS_{j+2}^zS_{j+3}^z\ket.
\label{ap4}
\end{align}
%
Here and hereafter $P_{\varepsilon_1\varepsilon_{2}
\varepsilon_{3}\varepsilon_{4}}^
{\tilde{\varepsilon}_{1}\tilde{\varepsilon}_{2}
\tilde{\varepsilon}_{3}\tilde{\varepsilon}_{4}}$ 
(also written as $P_{\varepsilon}^{\tilde{\varepsilon}}$ 
for simplicity) denotes a correlator of the form
%
\begin{equation}
P_{\varepsilon_1\varepsilon_{2}\varepsilon_{3}\varepsilon_{4}}^
{\tilde{\varepsilon}_{1}\tilde{\varepsilon}_{2}\tilde{\varepsilon}_{3}
\tilde{\varepsilon}_{4}}=\bra 
E_1^{\tilde{\varepsilon}_1\varepsilon_1}
E_{2}^{\tilde{\varepsilon}_{2}\varepsilon_{2}} 
E_{3}^{\tilde{\varepsilon}_{3}\varepsilon_{3}} 
E_{4}^{\tilde{\varepsilon}_{4}\varepsilon_{4}} \ket,
\label{P}
\end{equation}
%
where $\varepsilon_j$, $\tilde{\varepsilon}_j=\{+,-\}$ and 
$E_j^{\varepsilon_j\tilde{\varepsilon}_j}$ is the 2 $\times$ 2 
elementary matrix ($E^{\pm \pm}=\pm S^z+1/2$, $E^{- +}=S^{+}$, 
$E^{+-}=S^-$) acting on $j$th site. In this notation $P(4)=
P_{++++}^{++++}$. Note that because the 
Hamiltonian \eqref{Hami} has the symmetry under the group $SU(2)$, 
$P_{\varepsilon}^{\tilde{\varepsilon}}$ 
possesses a property like
%
\begin{equation}
P_{\varepsilon}^{\tilde{\varepsilon}}
=
P^{\varepsilon}_{\tilde{\varepsilon}}
=
P_{-\varepsilon}^{-\tilde{\varepsilon}}
=
P^{-\varepsilon}_{-\tilde{\varepsilon}}.
\label{psym}
\end{equation}
%
As in the case of $P(4)$, the correlators 
\eqref{P} enjoy 
the multiple integral representation
\cite{JMMN,JMbook,Naka,JM}:
%
\begin{equation}
P_{\varepsilon}^{\tilde{\varepsilon}}
=\prod_{j=1}^4\int_{C}\frac{d \lam_j}{2\pi i}
U(\lam_1,\dots,\lam_4)T(\lam_1,\dots,\lam_4),
\label{mir}
\end{equation}
%
where the integration contour $C$ is taken to be a line 
$[-\infty-i \alpha,\infty-i\alpha]$ $(0<\alpha <1)$.
For convenience,  we choose $\alpha=1/2$.
The integrand $U(\lam_1,\dots,\lam_4)$ is 
given by
%
\begin{equation}
U(\lam_1,\dots,\lam_4)=\pi^{10}\frac{\prod_{1\le k<j\le 4}
\sh\pi \lam_{jk}}{\prod_{j=1}^{4}\sh^4\pi \lam_j},
\label{ufnc}
\end{equation}
%
while $T(\lam_1,\dots,\lam_4)$ depends on the selection 
of $\varepsilon $ and $\tilde{\varepsilon}$.
Here and hereafter we use the notation 
 $\lam_{jk}=\lam_j-\lam_k$ to save space.
In particular for the correlator \eqref{ap4}, 
$T(\lam_1,\dots,\lam_4)$ is given by
%
\begin{equation}
T(\lam_1,\dots,\lam_4)=
\frac{\lam_1(\lam_1+i)^2\lam_2^3 \lam_3(\lam_3+i)^2 \lam_4^3}
{(\lam_{21}-i)\lam_{31}\lam_{41}\lam_{32}\lam_{42}(\lam_{43}-i)}.
\label{tfnc}
\end{equation}
%

To calculate the multiple integral \eqref{mir}, 
we follow the method 
by Boos and Korepin \cite{BK2}. Roughly, their method 
is described as follows. First Taking carefully into account the property 
of $U(\lam_1,\dots,\lam_4)$, we modify the integrand
$T(\lam_1,\dots,\lam_4)$ such that the integral gives
the same result as the original one (``weak equivalence").
In this way it is likely that the integrand 
$T(\lam_1,\dots,\lam_4)$ can always 
be reduced to the following form 
(we call it ``canonical form"):
%
\begin{align}
T_c=&
P_0(\lam_2,\lam_3,\lam_4)+
\frac{P_1(\lam_1,\lam_3,\lam_4)}{\lam_{21}}+
\frac{P_2(\lam_1,\lam_3)}{\lam_{21}\lam_{43}},
\label{canonical}
\end{align}
%
where $P_0$, $P_1$ and $P_2$ are
certain polynomials. 
Once one derives the  canonical form,
one can perform the multiple integral \eqref{mir} by using the Cauchy 
theorem \cite{BK2}. The result is 
written as combinations of the
logarithm $\ln 2$, the Riemann zeta function $\zeta(3)$ and 
$\zeta(5)$ and rational numbers. Consequently, the main part of the 
calculation for the multiple integral \eqref{mir} reduces 
to finding the canonical form \eqref{canonical}. 

Now we consider the  case \eqref{tfnc} and show
that the Boos--Korepin method is also applicable 
to our case. Let us introduce the following diagram. 
\[
\graphaa:=\frac{1}{\lam_{jk}},\qquad \graphbb:=\frac{1}{\lam_{jk}-i},
\]
where $j>k$.
First we 
expand \eqref{tfnc} through partial fractions. 
The result consists of 24 terms.
Taking into account the antisymmetry of the function 
\eqref{ufnc} under transposition of any two variables 
$\lam_j$ and $\lam_k$ and the symmetry of $T$ \eqref{tfnc}
under $\lam_1\leftrightarrow\lam_3$,
$\lam_2\leftrightarrow\lam_4$, we can reduce the 
24 partial fractions to eight ones.
Diagrammatically, its denominators are written as
\begin{align}
\grapha\to &\graphi-\graphb+i\left(\graphd+\graphh\right)\nn \\
+2i&\left(\graphf-\graphe-\graphg-\graphc\right).\nn
\end{align}
{}From a symmetry of the denominator, the second term 
can further be simplified as
\begin{align}
&-g\graphb
\to-\frac{1}{4}\left(g(\lam_1,\lam_3,\lam_2,\lam_4)-
g(\lam_4,\lam_1,\lam_3,\lam_2)\right. \nn \db
&\quad\left.+g(\lam_2,\lam_4,\lam_1,\lam_3)-
g(\lam_3,\lam_2,\lam_4,\lam_1)\right)\graphm \nn \\[4mm]
&\to\frac{1}{2}\lam_1\lam_2\lam_3\lam_4(i\lam_1+i\lam_3+2\lam_1\lam_3)
(i\lam_2+i\lam_4+2\lam_2\lam_4) \nn \\
&\quad\times\graphdd\times\graphee\, , \nn
\end{align}
where  $g$ denotes the numerator of \eqref{tfnc}.
Next using the Cauchy theorem, we shift variables  
$\lam_j\to\lam_j\pm i$ such that the denominators do not 
contain $i$. For instance, we have
\begin{align}
g \graphg&=g^{(1)} \graphj+g^{(2)}\graphk \nn \\[4.5mm] \nn
   &\to g^{(1)}\graphj+g^{(2)}_{\lam_2\to\lam_2+i}\graphl, \\ \nn
\end{align}
where
\begin{align}
g^{(1)}&=-\lam_1(\lam_1+i)\lam_2^3\lam_3(\lam_3+i)^2\lam_4^3 \nn \\
g^{(2)}&=-\lam_1(\lam_1+i)\lam_2^4\lam_3(\lam_3+i)^2\lam_4^3. \nn
\end{align}
Finally again using 
the antisymmetric property of \eqref{ufnc}, we eliminate 
the symmetric part with respect to $\lam_j\leftrightarrow\lam_k$.
In this way we have obtain the canonical form as
%
\begin{align}
P_0=&\frac{56}{5}\lam_2\lam_3^2\lam_4^3,\nn \db
P_1=&\frac{27}{10}\lam_4-i\lam_4^2+\frac{33}{5}\lam_3 \lam_4^2+
\frac{4}{5}\lam_4^3+2 i\lam_3\lam_4^3+4\lam_3^2\lam_4^3\nn \db
&+\lam_1(-4i\lam_4+7\lam_4^2-32i\lam_3\lam_4^2-10
i\lam_4^3-12i\lam_3^2\lam_4^3) \nn \db
&+\lam_1^2(4\lam_4-19i\lam_4^2-28\lam_3\lam_4^2-10\lam_4^3 \nn \db
&\qquad\qquad
\qquad\qquad
\qquad\quad-28i\lam_3\lam_4^3-32\lam_3^2\lam_4^3), \nn \db
P_2=&-\frac{3}{10}+\frac{3}{2}i \lam_3+\frac{3}{2}\lam_1\lam_3
-\frac{1}{2}\lam_3^2+4i\lam_1\lam_3^2+6\lam_1^2\lam_3^2. \nn
\end{align}
%
Subsequently, applying the method
developed in \cite{BK2}, we calculate the multiple integral
by substituting the above canonical
form into \eqref{mir}. Explicitly,
%
\begin{align}
&P_{+-+-}^{+-+-}=J_0+J_1+J_2 \nn \db
&J_0=\frac{7}{10}, \qquad 
J_1=-\frac{2}{3}+\frac{3}{10}\zeta(3)+\frac{35}{32}\zeta(5),\nn \db
&J_2=-\frac{1}{2}\zeta(3)+\frac{1}{2}\zeta(3)\ln2+\frac{9}{80}\zeta(3)^2
-\frac{25}{32}\zeta(5) \nn \db
&\qquad
-\frac{5}{8}\zeta(5)\ln2,
\label{integral}
\end{align}
where $J_k$ denotes the result of the integration 
regarding the term $P_k$.
We remark that the canonical form is not unique due to 
the non-uniqueness of partial fraction expansions. Accordingly, 
the explicit value of each $J_k$ depends on the choice of the 
canonical form. The final 
result $J_0+J_1+J_2$, however, is always unique as a matter of 
course.
%
%
%
%

Combining the result \eqref{integral} with \eqref{p4} 
and \eqref{ap4}, we obtain the third neighbor correlator 
$\bra S_j^{z}S_{j+3}^z \ket$ 
\eqref{3rd} and  at the same time the correlator 
$\bra S_{j}^z S_{j+1}^z S_{j+2}^z S_{j+3}^z\ket$ as
%
\begin{align}
\bra S_{j}^z S_{j+1}^z S_{j+2}^z S_{j+3}^z\ket&=
\frac{1}{80}-\frac{1}{3}\ln 2+\frac{29}{30}\zeta(3)-\frac{2}{3}\zeta(3)\ln 2
\nn \db
-\frac{21}{80}&\zeta(3)^2-\frac{95}{96}\zeta(5)+
\frac{35}{24}\zeta(5)\ln 2.
\label{4body}
\end{align}
%

Now let us consider the four-point correlators of
the form $\bra S_j^{\alpha} S_{j+1}^{\beta} S_{j+2}^{\gamma}
S_{j+3}^{\delta} \ket$ ($\{\alpha,\beta,\gamma,\delta\}\in 
\{x,y,z, 0\};\,S_j^0=\openone$).
Because the correlators with an odd number of $S^{\{\alpha\}\diagdown 0}$ 
vanish,  the possible non-zero correlators are restricted to 
the following three types:
$\bra S_j^{\alpha}S_{j+1}^{\alpha}S_{j+2}^{\beta}S_{j+3}^{\beta}\ket$,
$\bra S_j^{\alpha}S_{j+1}^{\beta}S_{j+2}^{\alpha}S_{j+3}^{\beta} \ket$
and
$\bra S_j^{\alpha}S_{j+1}^{\beta}S_{j+2}^{\beta}S_{j+3}^{\alpha}\ket$.
Further due to the isotropy of the Hamiltonian
\eqref{Hami},  one can find that the independent 
correlators are  written as the following seven ones:
$\bra S_{j}^x S_{j+1}^z S_{j+2}^z S_{j+3}^x\ket$,
$\bra S_{j}^z S_{j+1}^x S_{j+2}^z S_{j+3}^x\ket$,
$\bra S_{j}^z S_{j+1}^z S_{j+2}^x S_{j+3}^x\ket$
and already obtained ones in \eqref{1st}--\eqref{3rd} 
and \eqref{4body}. Then we shall calculate the remaining three correlators
here.
For convenience we use the operator 
$S^{\pm}(=S^x\pm i S^y)$ instead of $S^x$.
First we consider the  correlator 
$\bra S_j^{+}S_{j+1}^{z}S_{j+2}^z S_{j+3}^{-} \ket
(=2\bra S_j^{x}S_{j+1}^{z}S_{j+2}^z S_{j+3}^{x} \ket)$.
%
{}From \eqref{P} and the property \eqref{psym}, this correlator 
is expressed as
%
$\bra S_j^{+}S_{j+1}^{z}S_{j+2}^z S_{j+3}^{-} \ket
=\(P_{-+++}^{+++-}-P_{-+-+}^{++--}\)/2$.
%
Using the relation $\bra S_{j}^{+}S_{j+3}^{-}\ket=
2(P_{-+++}^{+++-}+P_{-+-+}^{++--})$, 
we obtain
%
\begin{equation}
\bra S_{j}^+ S_{j+1}^z S_{j+2}^z S_{j+3}^-\ket=
P_{-+++}^{+++-}-\frac{1}{2}\bra S_j^z S_{j+3}^z\ket.
\label{p2s1}
\end{equation}
%
Here we have used the relation $\bra S_{j}^{+}S_{j+3}^{-}
\ket=2\bra S_{j}^{z}S_{j+3}^{z}\ket$. 
Similarly the other correlators are given by
%
\begin{align}
\bra S_{j}^z S_{j+1}^+ S_{j+2}^z S_{j+3}^-\ket&=
P_{+-++}^{+++-}-\frac{1}{2}\bra S_j^z S_{j+2}^z\ket, \nn \db
\bra S_{j}^z S_{j+1}^z S_{j+2}^+ S_{j+3}^-\ket&=
P_{++-+}^{+++-}-\frac{1}{2}\bra S_j^z S_{j+1}^z\ket.
\label{p2s2}
\end{align}
%
Therefore our goal is to evaluate the auxiliary 
correlators $P_{-+++}^{+++-}$, $P_{+-++}^{+++-}$ and 
$P_{++-+}^{+++-}$.  They are given if we replace 
the integrand  $T(\lam_1,\dots,\lam_4)$ by
%
\[
T^{(l)}=
\frac{(\lam_1+i)^3\lam_2(\lam_2+i)^2(\lam_3+i)\lam_3^2
\lam_4^{4-l}(\lam_4+i)^{l-1}}
{(\lam_{21}-i)(\lam_{31}-i)
 (\lam_{32}-i)\lam_{41}\lam_{42}\lam_{43}}.
\]
%
in the multiple integral representation.
Here the correlator $P_{+++-}^{-+++}$, $P_{+++-}^{+-++}$ 
and $P_{+++-}^{++-+}$ correspond to $l=1$, 2 and 3, 
respectively. 
Using the procedure similar to the case of 
$P_{+-+-}^{+-+-}$,  one obtains the explicit
values of the above auxiliary correlators.
%
%
As a result, combining the identity \eqref{p2s1} and 
\eqref{p2s2} with \eqref{1st}--\eqref{3rd}, 
we arrive at
%
\begin{align}
\bra S_{j}^+ S_{j+1}^z S_{j+2}^z S_{j+3}^-\ket&=
\frac{1}{120}-\frac{1}{2}\ln 2+\frac{169}{120}\zeta(3)\nn \db
-\frac{5}{6}\zeta(3)\ln 2-&\frac{3}{10}\zeta(3)^2-\frac{65}{48}\zeta(5)+
\frac{5}{3}\zeta(5)\ln 2, \label{green2}\db
\bra S_{j}^+ S_{j+1}^z S_{j+2}^- S_{j+3}^z\ket&=
\frac{1}{120}-\frac{1}{3}\ln 2+\frac{77}{60}\zeta(3)\nn \db
-\frac{5}{6}\zeta(3)\ln 2-&\frac{3}{10}\zeta(3)^2-\frac{65}{48}\zeta(5)+
\frac{5}{3}\zeta(5)\ln 2,  \label{chiral}\db
\bra S_{j}^+ S_{j+1}^- S_{j+2}^z S_{j+3}^z\ket&=
\frac{1}{120}+\frac{1}{6}\ln 2-\frac{91}{120}\zeta(3)\nn \db
+\frac{1}{3}\zeta(3)\ln 2+\frac{3}{40}&
\zeta(3)^2+\frac{35}{48}\zeta(5)-\frac{5}{12}\zeta(5)\ln 2.
\end{align}
%

We mention a few remarks of our results.
(i) All the above correlators are written as
the logarithm $\ln 2$, the Riemann zeta function 
$\zeta(3)$ and $\zeta(5)$. This agrees with the 
general conjecture by Boos and Korepin: 
\textit{arbitrary correlators of the \xxx antiferromagnet 
are described as certain combinations of logarithm $\ln 2$, 
the Riemann zeta function with odd arguments and rational 
coefficients.}
Especially intriguing is the existence of the non-linear 
terms such as  $\zeta(3)^2$, $\zeta(3)\ln 2$ and 
$\zeta(5)\ln2$. 
{(ii)} The correlator \eqref{green2} is interpreted as the third 
neighbor one-particle Green functions $\bra c^{\dagger}_j 
c_{j+3} \ket_{\rm f}/4$ via the Jordan-Wigner transformation 
\eqref{JW}.
Obviously, the first neighbor one-particle Green
function is expressed as
\[ 
\bra c_j^{\dagger}c_{j+1}\ket_{\rm f}
=\frac{1}{6}-\frac{2}{3}\ln2\simeq-0.295431453707,
\] 
%
which coincides with $\bra S_j^+S_{j+1}^-\ket$. 
Due to the characteristic $\pi/2$-oscillation:
$\bra c^{\dagger}_j c_{k} \ket_{\rm f} \sim A_{jk}
\cos(\pi(k-j+1)/2)$, one finds 
$\bra c^{\dagger}_j c_{j+2k}\ket_{\rm f}=0$. 
Therefore the quantity \eqref{green}
is the first non-trivial exact result of the 
correlators containing the fermionic nature.
{(iii)} The difference between \eqref{green2} and 
\eqref{chiral} gives the nearest chiral correlator
\begin{align}
\bra (\bm{S}_{j}\times &\bm{S}_{j+1})
\cdot(\bm{S}_{j+2}\times \bm{S}_{j+3})\ket
=3(\bra S_{j}^+ S_{j+1}^z S_{j+2}^- S_{j+3}^z\ \ket 
\nn \db
&-\bra  S_{j}^+ S_{j+1}^z S_{j+2}^z S_{j+3}^-\ket)
=\frac{1}{2}\ln2-\frac{3}{8}\zeta(3), \nn
\end{align}
which exactly agrees with the one derived from the
ground state energy of an integrable  two-chain model with 
four-body interactions \cite{MT}.

To confirm the validity of our formulae, we performed numerical 
calculations by using the density-matrix renormalization group (DMRG) 
\cite{White92,White93} and numerical diagonalization.
As for the DMRG, we followed standard algorithm \cite{Peschel99}.
We have repeated renormalization 500-times.
At each renormalization, we kept, at most, 200
relevant states for a (new) block.
%
The numerical diagonalization was performed for the system size 
$L=24$, 28 and $32$. We extrapolate the data from a 
fitting function $a_0+a_1/L^2+a_2/L^4$. 
All our analytical results coincide quite accurately with 
both numerical ones (TABLE I).
%
\begin{table}
\caption{Estimates of the correlators by the
exact evaluations, DMRG and the extrapolations 
from the numerical diagonalization for the system size
$L=24, 28, 32$.}
\begin{ruledtabular}
\begin{tabular}{lrrr}
Correlators&Exact&DMRG&Extrap. \\
\hline
$\bra S_j^z S_{j+3}^z \ket$&-0.0502486&-0.0502426& -0.0502475 \\
$\bra S_{j}^+ S_{j+1}^z S_{j+2}^z S_{j+3}^-\ket$
\footnote{$\frac{1}{4}
\bra c_j^{\dagger}c_{j+3}\ket_{\rm f}$}&0.0205719&0.0205681
&  0.0205716\\
$\bra S_{j}^z S_{j+1}^z S_{j+2}^z S_{j+3}^z\ket$&0.0307153&0.0307105&
0.0307154\\
$\bra S_{j}^+ S_{j+1}^z S_{j+2}^- S_{j+3}^z\ket$&-0.0141607& -0.0141579
&-0.0141606\\
$\bra S_{j}^+ S_{j+1}^- S_{j+2}^z S_{j+3}^z\ket$&0.0550194&0.0550108
& 0.0550198
\end{tabular}
\end{ruledtabular}
\end{table}
%
%
%

In closing we would like to comment on generalizations 
of the present results.
The extension to the calculation of higher neighbor 
correlators $\bra S_j^z S_{j+k}^z\ket_{k\ge 4}$ is of great 
interest. The fourth neighbor one $\bra S_j^zS_{j+4}^z\ket$, 
for example, will be calculated by combination of the 
EFP, $P(5)$ and two independent auxiliary correlators, which can
in principle be evaluated. 
In fact $P(5)$ has been already obtained in \cite{BKNS}. 
The computation, however, will be much more complicated.  
Alternatively, extending the present 
result to the inhomogeneous case as in \cite{BKS} and
taking into account the property of the quantum Knizhnik-Zamolodchikow 
equation, we may derive higher neighbor correlators.
Using this, eventually we hope to 
extract the 
long-distance asymptotics $\bra S_j^{z}S_{j+k}^z
\ket _{k \gg 1}$, which is a crucial problem  
in conformal field theory \cite{Affleck,LT}.

The authors are grateful to H.E. Boos and V.E. Korepin
for many valuable discussions.
KS is supported by the JSPS research fellowships for 
young scientists. MS and YN are supported by Grant-in-Aid 
for young scientists No.~14740228 and No.~13740240, 
respectively.
This work is in part supported
by Grant-in Aid for the Scientific Research (B) No.~14340099
from the Ministry of Education, Culture, Sports, Science and 
Technology, Japan.

%


\begin{thebibliography}{}
%
\bibitem{Bethe} H.A.~Bethe, Z. Phys. {\bf 71}, 205 (1931).
%
\bibitem{hult} L.~Hulth\'{e}n, Ark. Mat. Astron. Fys. A {\bf 26}, 
1 (1938).
%
\bibitem{Taka}  M.~Takahashi, J. Phys. C {\bf 10}, 1289 (1977).
%
\bibitem{Takabook} M.~Takahashi, 
\textit{Thermodynamics of 
One-Dimensional Solvable Models}, (Cambridge University Press,
Cambridge, 1999).
%
%
\bibitem{DitIno} J.~Dittrich and V.I.~Inozemtsev, J. Phys. A {\bf 30},
L623 (1997).
%
\bibitem{JMMN} M.~Jimbo, K.~Miki, T.~Miwa, and, A.~Nakayashiki, 
Phys. Lett. A {\bf 168}, 256 (1992).
%

\bibitem{JMbook} M.~Jimbo and T.~Miwa, \textit{Algebraic Analysis of
Solvable Lattice Models}, (American Mathematical Society,
Providence, RI, 1995). 
%
\bibitem{Naka} A.~Nakayashiki, Int. J. Mod. Phys. A {\bf 9}, 5673 (1994).
%
\bibitem{KIEU} V.E.~Korepin, A.G.~Izergin, F.H.L.~E\ss ler, and D.B.~Uglov,
Phys. Lett. A {\bf 190}, 182 (1994).
%
\bibitem{JM} M.~Jimbo and T.~Miwa, J. Phys. A {\bf 29}, 2923 (1996).
%
\bibitem{KMT} N.~Kitanine, J.M.~Maillet, and V.~Terras,
Nucl. Phys. B {\bf 567}, 554 (2000).
%
\bibitem{Quano} Y-H.~Quano, J. Phys. A {\bf 35}, 9549 (2002).
%
\bibitem{BK1} H.E.~Boos and V.E.~Korepin, J. Phys. A {\bf 34}, 5311 (2001).
%
\bibitem{BK2} H.E.~Boos and V.E.~Korepin,
\textit{Integrable models and Beyond}, edited by 
 M.~Kashiwara and T.~Miwa (Birkh\"auser, Boston, 2002); hep-th/0105144.
%
%
\bibitem{BKNS} H.E.~Boos, V.E.~Korepin, Y.~Nishiyama, and M.~Shiroishi,
J. Phys. A {\bf 35} 4443 (2002).
%
\bibitem{BKS} H.E.~Boos, V.E.~Korepin, and F.A.~Smirnov,
 hep-th/0209246.
%
%
%
%
%
\bibitem{MT} N.~Muramoto and M.~Takahashi,
J. Phys. Soc. Jpn. {\bf 68}, 2098 (1999).
%
\bibitem{White92}S. R. White, Phys. Rev. Lett. {\bf 69}, 2963 (1992).
%
\bibitem{White93}S. R. White, Phys. Rev. B {\bf 48}, 10345 (1993). 
%
\bibitem{Peschel99}
{\it Density-Matrix Renormalization:
A New Numerical Method in Physics}, 
edited by
I. Peschel, X. Wang, M. Kaulke and K. Hallberg
(Springer-Verlag, Berlin, 1999).
%
\bibitem{Affleck} I. Affleck, J. Phys. A {\bf 31}, 4573 (1998) .
%
\bibitem{LT} S. Lukyanov and V. Terras, hep-th/0206093.
\end{thebibliography}
\end{document}